\documentclass[twocolumn,superscriptaddress,showpacs]{revtex4-1}
\usepackage{tabularx}
\usepackage{graphicx}
\usepackage{dcolumn}
\usepackage{bm}
\usepackage{xcolor}
\usepackage{color}
\usepackage{amsmath,amsfonts,amssymb}
\newcommand{\svo}{Sr$_2$VO$_4$}

\begin{document}

\title{Jahn-Teller induced nematic orbital order in tetragonal Sr$_2$VO$_4$}

\author{J. Teyssier}
\affiliation{Department of Quantum Matter Physics, University of Geneva, Quai Ernest-Ansermet 24, 1211 Gen\`eve 4, Switzerland}

\author{E. Giannini}
\affiliation{Department of Quantum Matter Physics, University of Geneva, Quai Ernest-Ansermet 24, 1211 Gen\`eve 4, Switzerland}

\author{A. Stucky}
\affiliation{Department of Quantum Matter Physics, University of Geneva, Quai Ernest-Ansermet 24, 1211 Gen\`eve 4, Switzerland}

\author{R. \u{C}ern\'y}
\affiliation{Department of Quantum Matter Physics, University of Geneva, Quai Ernest-Ansermet 24, 1211 Gen\`eve 4, Switzerland}

\author{M.~V.~Eremin}
\affiliation{Kazan (Volga region) Federal University, 420008 Kazan,
Russia}

\author{D. van der Marel}

\affiliation{Department of Quantum Matter Physics, University of Geneva, Quai Ernest-Ansermet 24, 1211 Gen\`eve 4, Switzerland}

\date{\today}

\begin{abstract}

Using high resolution X-Ray diffraction (XRD) on high purity powders, we resolved the structure and $ab$ symmetry of the intriguing compound \svo$ $
from room temperature down to 20 K to an unprecedented level of accuracy. Upon cooling, this new set of data unambiguously reveals a second order phase transition
lowering the symmetry from tetragonal to orthorhombic at a
temperature $T_{c2}=136$ K. The observation of an orthorhombic
distortion of the $ab$-plane is attributed to nematic phase
formation supported by local Jahn-Teller (JT) dynamical
instability. At $T_{N}=105$ K, spins order and at $T_{c1}=100$ K the tetragonal structure
is recovered with an elongated c-axis.
\end{abstract}
\maketitle
\section{Introduction}
\svo $ $ can be stabilized with vanadium ions in a $4^{+}$ valence
state in two different polytypes: (i) an orthorhombic form
whose peculiar magnetic dimer behavior has been previously
investigated \cite{Deisenhofer_Electron_2012} and (ii) a
tetragonal one that exhibits puzzling structural and magnetic transitions and
is the object of the present study. The ground state of this
Mott-Hubbard insulator has been studied experimentally
\cite{Cyrot_Properties_1990,Zhou_orbital-ordering_2007,Teyssier_Optical_2011,Sugiyama_Hidden_2014}
and theoretically
\cite{Pickett_Theoretical_1989,Imai_Electronic_2005,Jackeli_Magnetically_2009,Eremin_alternating_2011}
for a long time. Nevertheless, doubts remain about the nature
of the order parameter. Upon cooling, a magnetically ordered
ground state is achieved after a succession of structural and magnetic
transitions, whose nature is not yet understood. Specific heat
and magnetic susceptibility measurements have revealed the following magnetic
and structural diagram:
\begin{itemize}
\item{Below 10 K the material exhibits a tetragonal structure with a weak magnetic order \cite{Pickett_Theoretical_1989,Sugiyama_Hidden_2014}.}
\item{Between 10 K and 100 K the system enters a different magnetic and orbital state, which is not ferromagnetic and the nature of which has not
 yet been fully understood.}
\item{In the region between $T_{c1}\sim 100$ K and $T_{c2}\sim 140$ K recent structural analysis suggests an orthorhombic distortion of the $ab$-plane
  \cite{PhysRevB.92.064408} but the magnetic state remains not known.}
\item{Above $T_{c2}$ the material is again tetragonal with a slightly reduced $c$ axis \cite{Zhou_orbital-ordering_2007} and it exhibits a paramagnetic behaviour with a large Van Vleck contribution \cite{Eremin_alternating_2011}.}
\end{itemize}

The main candidates for the ground state that have been proposed until now are an octupolar ground state \cite{Jackeli_Magnetically_2009} and anti-ferro orbital ordering with muted magnetic moments due to spin-orbit coupling \cite{Teyssier_Optical_2011}.
These two scenarios are well supported by experiments for most of the temperature range between 10K and ambient temperature. Only the splitting of the excited state measured with inelastic neutron scattering \cite{PhysRevB.81.212401}, while consistent with anti-ferro orbital ordering \cite{Teyssier_Optical_2011}, is  inconsistent with octupolar order\cite{Jackeli_Magnetically_2009}. However, neither of these two models captures the two-stage ordering implied by the two peaks in the specific heat at 98 K and 127K. 
Two possible configurations account for the observed double structural transition at $T_{c1}$ and $T_{c2}$: (i) the coexistence of the low (LT) and the high temperature (HT) tetragonal phases, (ii) the onset of an intermediate orthorhombic distortion of the lattice associated to the transition from the HT to LT phase.

This ambiguity was recently removed by Yamauchi et al. \cite{PhysRevB.92.064408}, who concluded, based on Le Bail fitting of X-ray powder diffraction data, that the intermediate phase is orthorhombic. They suggested the space group $Immm$ as being the most likely, according to the group-subgroup relation to the $I$4$/mmm$ space group of the tetragonal phase. Nevertheless, the structural model of the orthorhombic phase was not resolved in Ref.\onlinecite{PhysRevB.92.064408}, and higher resolution diffraction data are needed for that purpose.

Here we present high resolution synchrotron X-ray diffraction data (XRD), and resolve the structural model of the orthorhombic phase that successfully fits the data, through a full pattern profile refinement performed using the Rietveld method.  We also present theoretical arguments that the orbital ordering in \svo is accompanied by an intermediate
orthorhombic distortion stabilized by electron-lattice
interactions. Examples of lattice mediated orbital ordering
are well known in KCuF$_3$ \cite{Lee_two-stage_2012} or in
LaMnO$_3$ \cite{Ahmed_Volume_2009,trokiner_melting_2013}. However,
in those compounds $e_g$ electrons are involved, therefore the
spin-orbit coupling does not play an essential role,
contrary to the situation in \svo.

\section{X-ray powder diffraction data}
Synchrotron X-ray powder diffraction experiments were carried
out at the high resolution diffraction beamline I11 at the
Diamond Light Source (details on experimental procedure and
crystallographic data analysis are given in appendix, details of the beamiline are reported in  \cite{Parker:he5503,ISI:000268615700046}). The samples were
prepared by reduction of Sr$_4$V$_2$O$_9$ in a sealed quartz tube 
with zirconium as an oxygen getter
\cite{1742-6596-200-1-012219}. No secondary phases nor
traces of the orthorhombic phase were detected after 5 consecutive
reduction cycles at 850 $^o$C for 48h.

\begin{figure}[htp!]
  \includegraphics[width=8.5 cm]{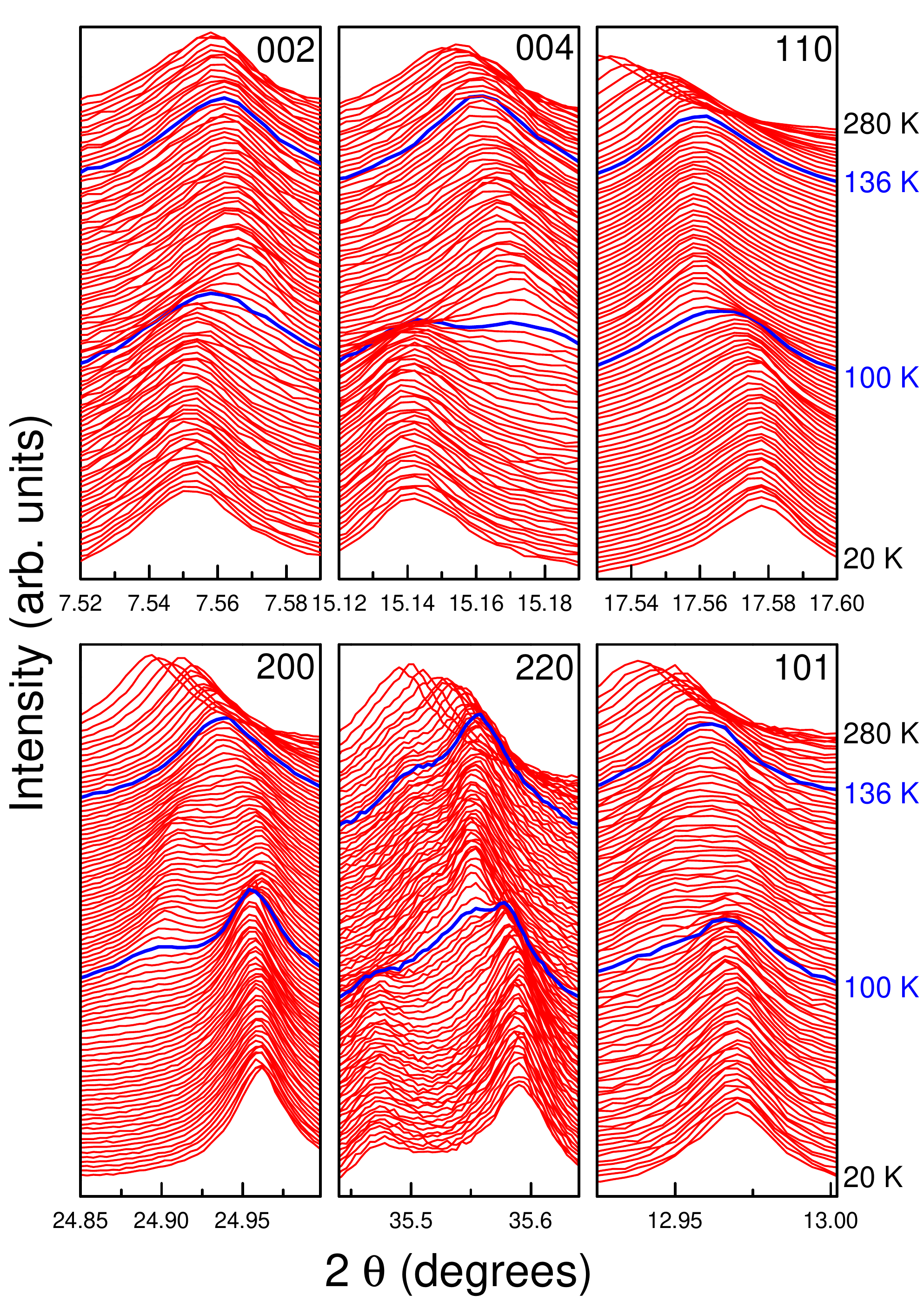}\\
  \caption{ a) Temperature evolution of selected Bragg reflections. The two transition temperatures $T_{c1}$ and $T_{c2}$ are highlighted with blue lines. Temperature intervals are 1 Kelvin from 20K to 40K, 5K from 45K to 85K, 1 Kelvin from 90K to 139K, 10K from 140K to 190K and 20K from 200K to 280K.}\label{peaks}
\end{figure}
\begin{figure}[t!]
  \includegraphics[width=8.5 cm]{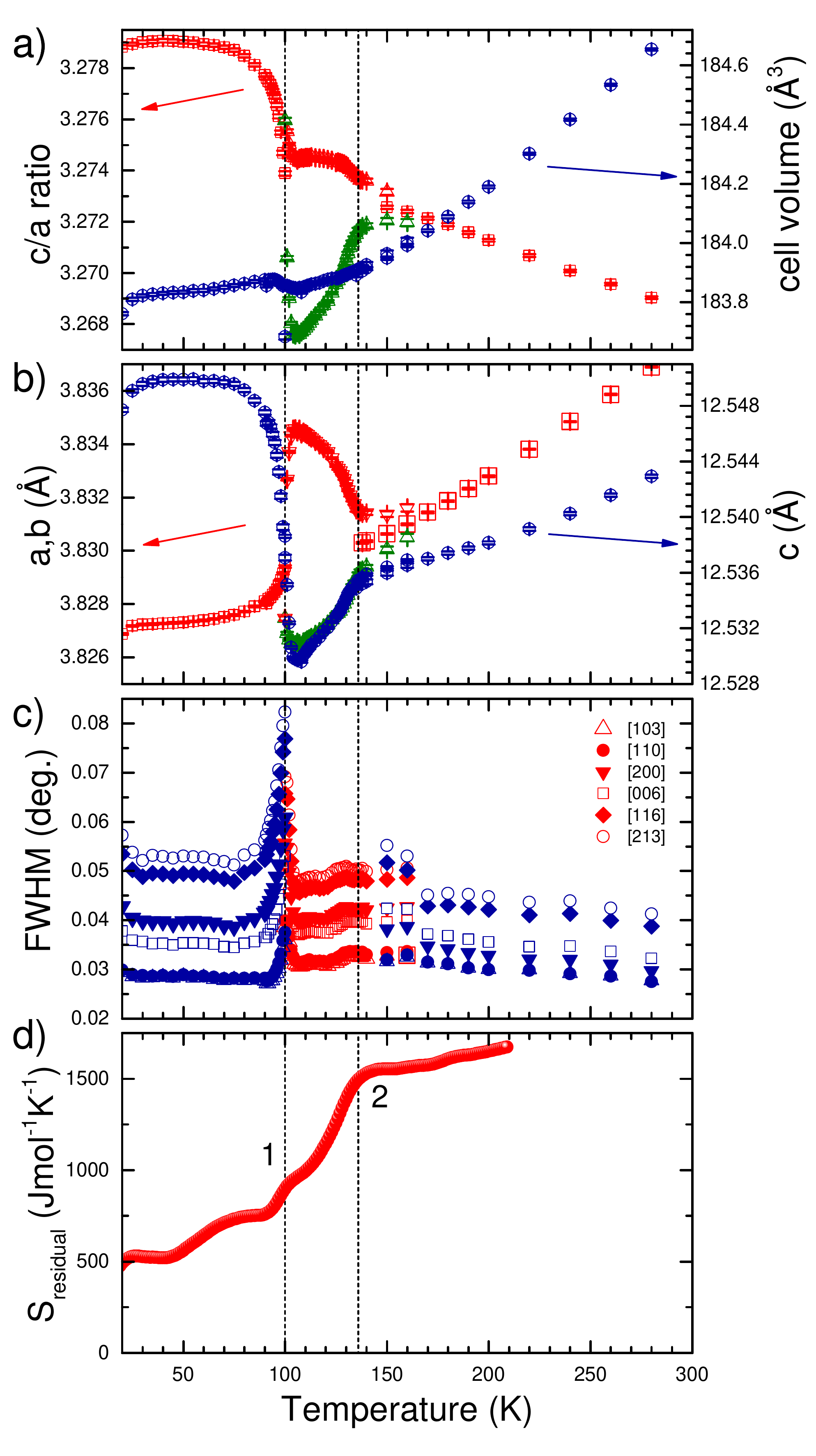}\\
  \caption{Temperature evolution of  a)  $c/a,b$ ratio and volume of the cell.  b) lattice parameters and c) Full Width Half Maximum (FWHM) of selected diffraction peaks in the high temperature and low temperature tetragonal phases (blue symbols) and in the orthorhombic phase (red and green symbols). The two transition temperatures $T_{c1}$ and $T_{c2}$ are highlighted with dotted lines. Note that error bars in a) and b) are smaller than the symbols. d) entropy corrected for the phonon and thermal expension (see methods in appendix)} \label{XRD}
\end{figure}

The temperature evolution of diffraction peaks in the
$ab$-plane (200 in Fig. \ref{peaks}) and along the $c$-axis
(002 and 004 in Fig. \ref{peaks}) clearly indicates a long
range $ab$-plane distortion through the splitting of the $200$
reflection between $T_{c1}$ and $T_{c2}$,  and a first-order like
$c$-axis expansion below $T_{c1}$. During the Rietveld refinement
(see details in appendix) the symmetry of the structure was set to tetragonal in the high
(above $T_{c2}$) and low (below  $T_{c1}$) temperature ranges which was
sufficient to index all diffraction peaks. The diffraction pattern
at 108 K, showing large departure from the two
tetragonal phases (Fig. \ref{XRD}b) and slightly above $T_N=105$ K, was used  to reliably identify the symmetry of the intermediate phase. A
dichotomy indexing routine implemented in the program \emph{Fox}
\cite{favre-nicolin_<i>fox</i>_2002} was used to index 20 low
angle peaks belonging to \svo. An orthorhombic cell corresponding
to the deformation of the low temperature tetragonal cell
($I$4$/mmm$) was immediately found, in agreement with a recent study in which long range orthorhombic distortion was
also suggested \cite{PhysRevB.92.064408}. The observed
extinctions pointed only to a body centered lattice which is in
agreement with 4 possible space groups. The highest symmetry space group among them,  $Immm$, was successfully used to solve the structure ab-initio by the global optimization method using the program \emph{Fox}, and modeling the structure by one Sr, one V and three O atoms. The resulting structural model corresponds to a deformation of the tetragonal structure with the VO$_6$ octahedron elongated along the $c$-axis, and slightly deformed in the equatorial plane (the parameters are provided in the appendix), which is in agreement with the loss of the 4-fold axis and resulting orthorhombic deformation. 

Upon cooling down to the transition at $T_{c2}$, observed also in the entropy extracted from specific heat (see methods in appendix and point 1 in Fig. \ref{XRD}d),  the $a$ and $c$ lattice parameters contract (Fig. \ref{XRD}b). 
Upon lowering the temperature below $T_{c2}$ the $a$ and $b$-axis parameters separate,
and the $c$-axis shrinks. The volume remains unchanged  at the transition at $T_{c2}$ (1\% of volume collapse was reported in
LaMnO$_3$ at the Jahn-Teller transition
\cite{chatterji_volume_2003,Ahmed_Volume_2009,trokiner_melting_2013}),
but it slightly expands (0.02\%) upon cooling below  $T_{c1}$
(Fig. \ref{XRD}a). The weak lattice contraction visible at the lowest temperatures is in agreement with the proximity of a transition to a magnetic ordered state \cite{Sugiyama_Hidden_2014}.
The evolution of the full width at half maximum (FWHM) of selected diffraction peaks is shown in Fig. \ref{XRD}c. No significant broadening is observed at the orbital ordering temperature at $T_{c2}$ pointing toward a second order like transition. A discontinuity is visible around $T_{c2}$ in which the pattern was refined to two separate structural models (orthorhombic, red-green symbols, and tetragonal, blue symbols, in Fig. \ref{XRD}c, too close to each other to take into account a likely coexistence of both of them.

\section{Discussion and interpretation}
\begin{figure}[h!]
  \includegraphics[width=8.5 cm]{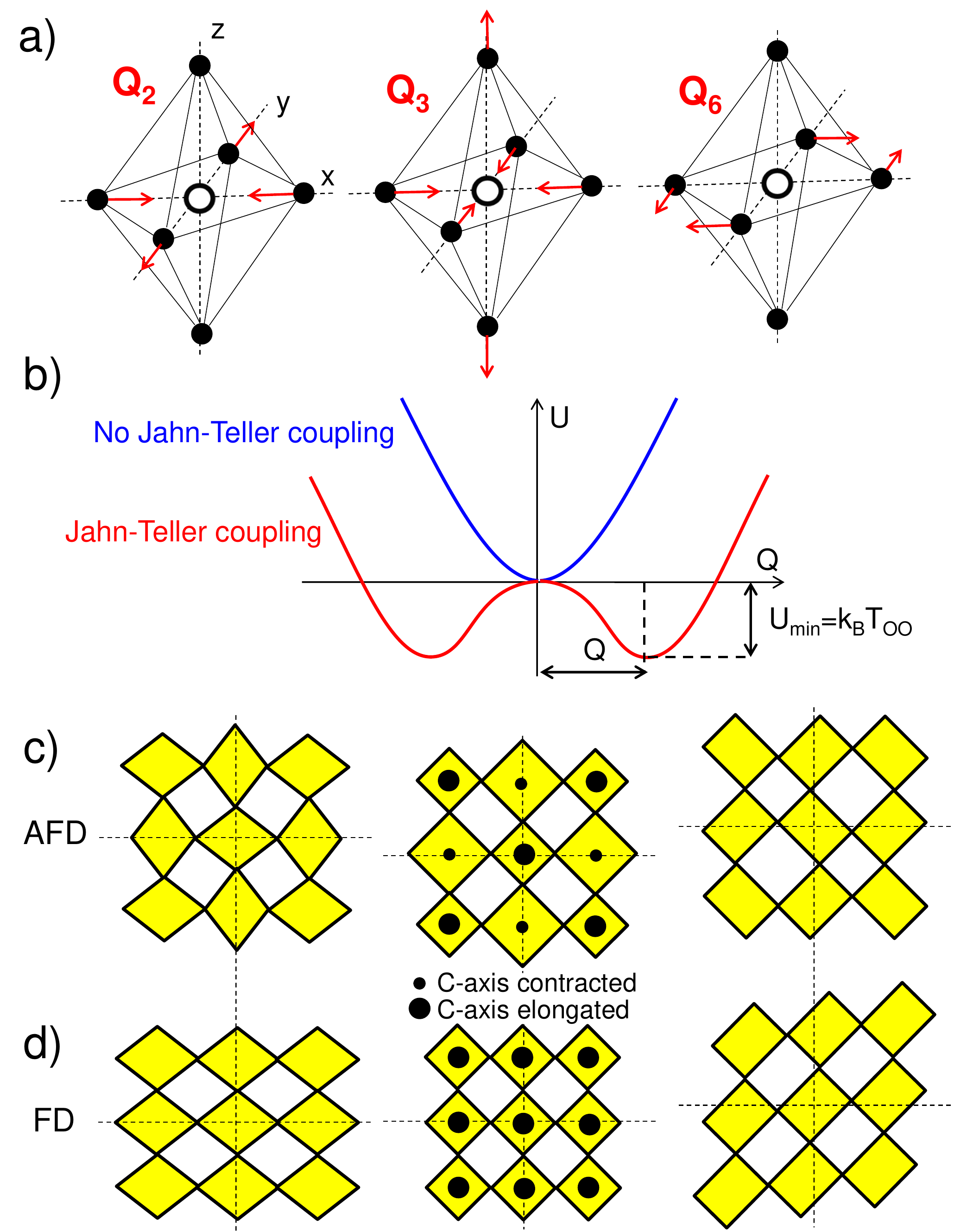}\\
  \caption{ a) representation of ionic displacement corresponding to the $Q_2$, $Q_3$ and $Q_6$ Jahn-Teller distortion modes. b) Scheme of the adiabatic potential profile due to electron lattice interaction. $Q$ is the distortion strength and $U_{min}=k_BT_{OO}$ the potential barrier.  Scheme of c) antiferro- and d) ferro-distortions associated to coupling with each mode.}\label{orbital}
\end{figure}

\begin{figure}[h!]
  \includegraphics[width=8.5 cm]{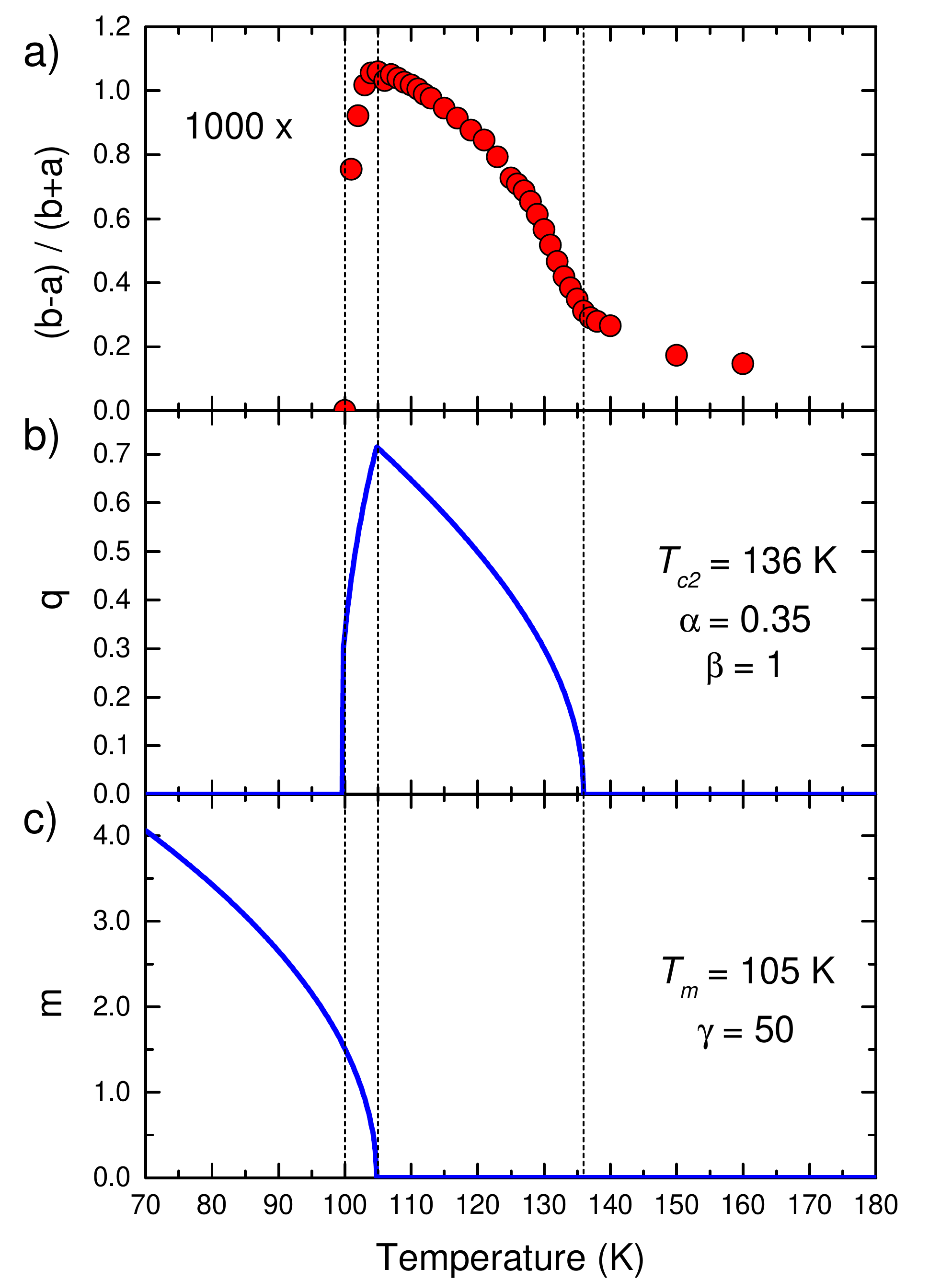}\\
  \caption{a) Experimental data of the Jahn-Teller order parameter $(a-b)/(a+b)$, b) calculation of the coupled Jahn-Teller order parameter $q$  and c) the anti-ferromagnetic order parameter $m$ using Eq. \ref{gl1}. The parameters are indicated in the figure and caracteristic temperatures $T_{c1}$, $T_N$ and $T_{c2}$  are marked as vertical dashed lines.  }\label{order}
\end{figure}

The XRD data indicates a local in-plane distortion which is not compatible with the orthorhombic distortion of the octahedra typically observed in KCuF$_3$ \cite{Lee_two-stage_2012} and LaMnO$_3$ \cite{chatterji_volume_2003}, where the 4 fold axis is preserved although each octahedron presents an orthorhombic distortion of its $ab$-plane.
Indeed, the tiny $ab$ splitting observed in our XRD measurements (Fig.\ref{order}) is only compatible with a breaking of $C4$ symmetry, {\em i.e.} a long range orthorhombic distortion (see also ferro-distortion in Fig. \ref{orbital}d). 

We propose the following model: Due to spin-orbit coupling each V-ion has a spin-orbital state in which spin ($s=1/2$) and orbital (effectively $l=1$) moments combine to a $j=1/2$ spin-orbital doublet with a muted magnetic moment due to the factor of 2 difference in spin and orbital gyromagnetic ratio. 
Due to the absence of long-range order of the muted-moments above $T_{c2}$, the crystal structure is tetragonal. 
Below the orbital ordering temperature ($T_{c2}$) coupling to Jahn-Teller (electron-lattice) modes (Fig. \ref{orbital}a) splits the adiabatic potential (Fig. \ref{orbital}b) driving the system to a local distortion. 
The locally distorted ions form a nematic phase. At the N\'eel temperature $T_N$=105 K antiferromagnetic ordering sets in.
The Jahn-Teller coupling switches off at $T_{c1}$, thus allowing the system to come back to a bipartite square lattice. 
This restores the tetragonal symmetry of the lattice while theoretically we expect a doubling of the unit cell.

In a previous publication \cite{Eremin_alternating_2011}, we have discussed the crystal field Hamiltonian including spin-orbit coupling and super-exchange interactions. The relevant orbital degrees of freedom is the set of degenerate vanadium 3d orbitals $3d_{xz}$ and $3d_{yz}$ for both spin-quantum numbers, which are mixed by spin-orbit interaction. These orbitals are coupled to the lattice degrees of freedom, $Q_j$ with $j=1,2,3,6$ of which $j=2,3,6$, depicted in Fig. \ref{orbital}a), are Jahn-Teller active. 
Only the $Q_2$ mode is sufficiently strongly coupled to be relevant in the present discussion. 
The adiabatic potential in reduced dimensionless units describing these couplings is
\begin{equation}\label{potential}
u(q)= \frac{q^2}{2}- v(T,m)\sqrt{1+q^2} 
\end{equation}
where $v(T,m)$ is the electron-lattice coupling constant normalized by the spin-orbit coupling constant, which varies as a function of temperature as well as the anti-ferromagnetic order parameter $m$ (see Appendix).  In particular $v>1$ for $T_{c1}<T<T_{c2}$ and $v<1$ above and below this temperature range. 
On the level of the individual $V^{4+}$-ions, the displacements of the surrounding oxygen ions are subject to zero-point motion. 
However, these motions are mechanically coupled between neighboring sites. The coordinate $Q_2$ describing the collective vibration of all ions then represents a macroscopic observable, subject to the occurrence of spontaneous symmetry breaking with two energetically equivalent minima for positive and negative $Q_2$ (Fig. 3b). 
The symmetry breaking disappears for high enough temperatures where thermal motion of the lattice overcomes the symmetry-broken state. 
The scenario is therefor as follows: For $T>T_{c2}$ the electrons occupy orbital momentum eigenstates $d_{xz,\sigma}\pm i d_{yz,\sigma}$, without long-range intersite magnetic order. 
Below $T_{c2}$ a collective Jahn-Teller effect breaks the electronic symmetry by shifting the orbital character towards $d_{xz,\sigma}$ or $d_{yz,\sigma}$ while at the same time moving $Q_2$ away from the origin. 
A similar dynamical regime is present in LaMnO$_3$ but at much higher temperature ($T_{JT} = 700$ K) due to strong electron-lattice coupling \cite{Ahmed_Volume_2009}. 
The static local Jahn-Teller effect usually leads to an anti-ferro distorted order, which globally preserves the tetragonal (C4) symmetry  (see Fig. \ref{orbital}c), but would induce a doubling of the unit cell in $ab$-plane.  
This is not compatible with our experimental observations where a long range distortion is observed (Fig. \ref{XRD} and Fig. \ref{orbital}d).  
We therefor conjecture that the dynamical Jahn-Teller local distortions enable a ferro-nematic long range order induced by interaction via the phonon field. 

The magnetic transition in Sr$_2$VO$_4$ occurs near 100 K. Until now the experimental data have been interpreted as a first order transition, {\em i.e.}  $m(T)$ jumps to a finite value at $T_{c1}$ (see Fig. 1 in ref.\onlinecite{Teyssier_Optical_2011}).  
The spin-orbit coupling causes the orbitals to polarize  as $d_{xz,\sigma}\pm i d_{yz,\sigma}$, and the Jahn-Teller instability is suppressed due to the fact that the parameter $v$ jumps to a value smaller than 1 (see Appendix). 
Indeed the tetragonal symmetry is recovered below $T_{c1}$ as observed in the crystallographic data in Fig. \ref{XRD}.
However, the evolution of the $a$-axis and $c$-axis parameters below $T_{c1}$  suggests another scenario where there is a second order anti-ferromagnetic transition at $T_N$. The magnetic order parameter $m(T)$ than increases continuously (Fig. \ref{order}c). 
At $T_{c1}$  the condition $v=1$ is reached, below which the Jahn-Teller distortion disappears. 
In this scenario there are three critical temperatures: $T_{c1}=100$ K, $T_N=105$ K and $T_{c2}=T_{JT}=136$ K.
The steps $T_{c1}$ K (point 1 in Fig. \ref{XRD}d) and $T_{c2}$ (point 2 in Fig. \ref{XRD}d) in the entropy are associated with the tetragonal to orthorhombic structural transitions that are visible in the crystallographic data around $T_{c1}$ K and $T_{c2}$. 
The temperature dependence of the Jahn-Teller distortion and the anti-ferromagnetic order are described by combining the Jahn-Teller adiabatic potential and a Ginzburg-Landau expression for the free energy of the anti-ferromagnetic order parameter, as detailed in the Appendix. The model calculation of the Jahn-Teller order parameter $q$ (Fig. \ref{order} b) and the experimental $a/b$ axis ratio (Fig. \ref{order} a) show qualitatively the same features:  A gradual rise of $q$ below $T_{c2}$, sudden break at $T_N$ and disappearance at $T_{c1}$. These similarities suggest that the observed temperature dependence of the $a/b$ splitting is described by the combined effect of a Jahn-Teller distortion between $T_{c1}$ and $T_{c2}$ and the anti-ferromagnetic order of the spin-orbital mute moments below $T_N$.

It is worth noting that a very similar scenario of interplay between nematic order and a small distortion of the crystal lattice was recently discussed in iron-based superconducting pnictides.
The instability in pnictides is not related to JT dynamical effect but to Fermi surface nesting \cite{fernandes_what_2014}.
The resulting nematic phase induces a very similar long range orthorhombic distortion with a re-entrant tetragonal structure at low temperature \cite{Avci_Magnetically_2014}.
A recent study on LaSrVO$_4$ reports that local Jahn-Teller distortions are responsible for a liquid orbital state \cite{dun_lasrvo4:_2014}.
This is indeed one of the signposts of the nematic order suggested by the present work.
The splitting in the crystal field optical excitation spectrum at $T_{c2}$ (Fig. 5 in Ref. \onlinecite{Teyssier_Optical_2011}), originally attributed to the coexistence of two tetragonal phases, probably arises from the lifting of the degeneracy of the lowest Kramers doublet by the lowering of the symmetry below the JT transition.\\

\section{Conclusions}
In conclusion our high resolution powder diffraction measurements on tetragonal \svo have revealed a second order transition to a weak orthorhombic distorted structure below $T_{c2}=136$ K. 
In the temperature window from  $T_{c2}$ down to $T_{c1}=100$ K, superexchange and spin-orbit coupling, which tend to favor an anti-ferro-orbital ground state, compete with a nematic orbital ordering supported by the dynamical Jahn-Teller effect.
Below the spin-orbital ordering temperature ($T_N=105$ K), super-exchange and spin-orbit interaction screen the Jahn-Teller effect and at $ T_{c1}$ the tetragonal symmetry is recovered.\\

\section{Acknowledgements}
We thank the Diamond Light Source for access to beamline I11 (EE7637) that contributed to the results presented here. 
This work is supported by the SNSF through grant No. 200021-146586. M.V.E. was partially funded by the subsidy allocated to Kazan Federal University for the state assignment in the sphere of scientific activities. 
We acknowledge partial support by the DFG via the Collaborative Research Center TRR 80.

\section{Appendix}
\subsection{Experimental methods}
\begin{figure}[h!]
  \includegraphics[width=8.5 cm]{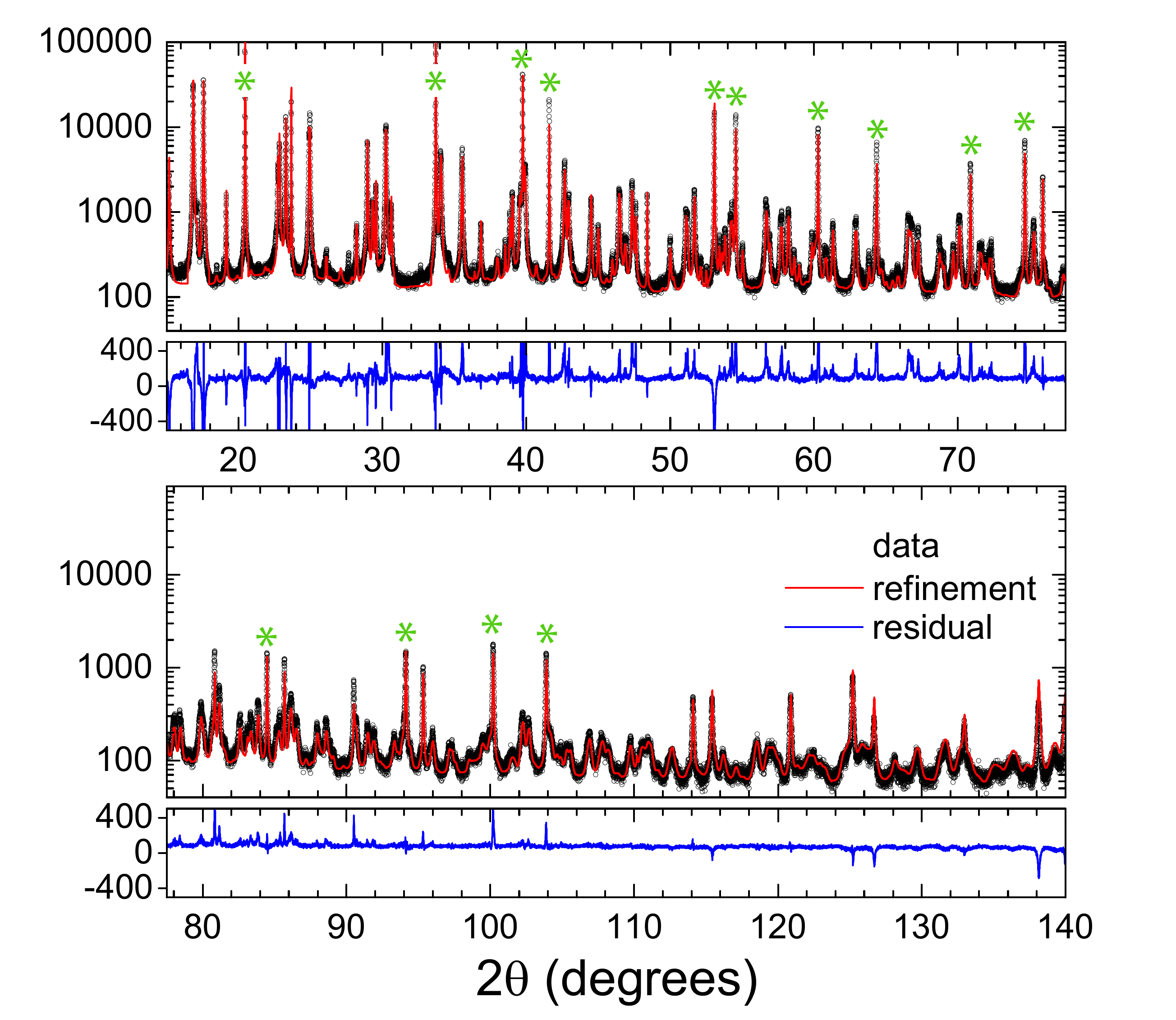}\\
  \caption{Rietveld refinement of the orthorhombic phase at 110 K. Diffraction peaks for alluminium are marked with green stars. }\label{refinement}
\end{figure}
Structural studies were performed through X-ray diffraction experiments on powder samples. Measurements were done at synchrotron facility at Diamond Light Source, by using the I11 beamline  \cite{Parker:he5503}. The X-ray beam was monochromatized at $E=15$ keV ($\lambda=0.82713$ \AA). The high-resolution detector was made of five multiple-analyzing-crystals and spanned a $2\theta$ range from $0.02^o$ to $150.00^o$ with a $\Delta \theta$ step of $0.004^o$ \cite{ISI:000268615700046}. The sample was made of a finely ground \svo  powder, homogeneously coating a thin Al wire perpendicular to the beam direction and the scattering vector. Thanks to this sample holder, any preferred orientation of \svo  grains was checked to be absent. The sample holder was mounted inside a cryogen-free refrigerator and the temperature was controlled in the range from $20$ to $300$ K, while XRD diffraction was acquired at constant temperature. Small temperature steps ($1$ K) were chosen over the range where structural transitions were expected to occur ($20-40$ K and $90-136$ K). With the chosen number of T steps, the acquisition time at each T and the heating rate ($0.1-1$ K/min), the whole diffraction experiment took about 25 hours. During the allocated beamtime, we could repeat the experiment twice, on two different powder samples, for checking the reproducibility.

The diffraction data analysis must carefully take into account the contribution of the sample holder which gives the strongest contribution to scattering. Nevertheless, Fig. \ref{refinement} shows a very strong signal to noise ratio of low-intensity diffraction peaks at large angle, which proves the high quality of the experimental data and allows extracting reliable information.

\begin{table}[]
\vspace{0.5cm}
\begin{tabular}{|c|c|}
\hline
parameter& T=280 K\\
\hline
a ($\AA$)&3.83689(2)\\
c ($\AA$)&12.54294(12)\\
\hline
$R_p$&0.164\\
\hline
$R_{wp}$&0.242\\
\hline
\end{tabular}\\
\vspace{0.5cm}
\begin{tabular}{|c|c|}
\hline
parameter& T=108 K\\
\hline
a ($\AA$)&3.82642(3)\\
b ($\AA$)&3.83438(3)\\
c ($\AA$)&12.52961(8)\\
\hline
$R_p$&0.155\\
\hline
$R_{wp}$&0.199\\
\hline
\end{tabular}\\
\vspace{0.5cm}
\begin{tabular}{|c|c|}
\hline
parameter& T=20 K\\
\hline
a ($\AA$)&3.82688(3)\\
c ($\AA$)&12.54771(16)\\
\hline
$R_p$&0.277\\
\hline
$R_{wp}$&0.368\\
\hline
\end{tabular}\\
\caption{lattice parameters and R-factors for the 3 crystallographic structures at 280K (high temperature (HT) tetragonal phase), 108 K (distorded orthorhombic phase), 20K (low temperature (LT) tetragonal phase). }\label{param}
\end{table}
A full pattern profile refinement based on the Rietveld method was used by means of the FullProf Suite program 2.05 \cite{RodriguezCarvajal199355}. The first parameters to be refined were those concerning the Al sample holder. As a matter of fact, the Al wire was found to have a filament-like (extrusion) texture, and therefore to exhibit a strong preferred orientation. 

Three known phases contribute to the diffraction pattern and were included in the pattern profile refinement: Al (the sample holder), \svo (either tetragonal or orthorhombic depending on the temperature range, as explained in the following), and Sr$_3$V$_2$O$_8$ (the only possible impurity phase present in the starting powder). 26 parameters were refined in total, namely the scale factors, lattice parameters, peak shape (pseudo-Voigt) for Al and \svo, preferred orientation and asymmetry correction for Al only, and zeroshift. After having refined the lattice parameters of all phases, the coordinate of the apical oxygen of the VO$_6$ octahedron was refined as well. The background was drawn by linear interpolation between manually selected points.

The profile refinement strategy was the following: in a first step we refined the scale, lattice, texture and asymmetry parameters of Al, and the zeroshift. Then, with all these parameters constrained to their convergence values, we refined scale and lattice parameters of the oxides. In a third step we refined lattice and shape parameters of \svo only, with all the others constrained at their primary convergence values, before a last step in which all parameters were free and refined together to the final convergence values. 

The R factors are reported to be in the range 0.15 - 0.4 (see Table I) and could seem to be rather high at first glance. The quality of the fits is mainly limited by i) the strong filament-like texture of the Al sample holder; ii) the impossibility to refine two (likely) coexisting tetragonal and orthorhombic \svo phases near to the transitions temperatures; iii) the fixed composition, and therefore the site occupation and position, of all phases; iv) the fixed zeroshift parameter over the whole temperature range.
Despite this affects the R-factor values, it has no influence on the lattice constants extracted from the structural refinements. 

The lattice parameters of Al were found to be in excellent agreement with the expected calibration curve. The secondary phase, Sr$_3$V$_2$O$_8$, was found to be present at $1.5-2\%$ volume fraction, in agreement with the preliminary laboratory XRD experiments that could not identify any secondary phases within the limit of instrument accuracy. 
Three different temperature ranges were chosen, in which three different structure models for \svo were used for the pattern refinement.
Below $T_{c1}$, and above $T_{c2}$, a tetragonal structure with space group $I$4$/mmm$ was chosen, whereas an orthorhombic one with a space group $Immm$ was used in the range $T_{c1}$-$T_{c2}$. The orthorhombic distortion at intermediate temperature was obtained from {\em ab-initio} structure solution. At crossover temperatures, corresponding to the structural transitions, each \svo phase was kept well beyond its stability limit in order to check the continuous overlap of the lattice parameters. Correspondingly, a divergence of the peak shape parameters (determining the full width at half maximum, FWHM) of the tetragonal and orthorhombic phases was found, as expected (see Fig. 2 c).

A summary of the lattice parameters is given in Table \ref{param} for the 3 crystallographic structures at 280K (High temperature (HT) tetragonal phase), 105 K (most distorted orthorhombic phase), 20K (Low temperature (LT) tetragonal phase)

The entropy shown in Fig. 2 d is the integration of residual specific heat after subtraction of the phonon contributions which we modeled with one Einstein and one Debye mode, and a linear term representing thermal expansion. 
\subsection{Details of the model}
Due to the tetragonal crystal field the Hilbert space relevant for the low energy properties is spanned by the degenerate set of single-electron states $\{|yz,+\rangle, |yz,-\rangle, |xz,+\rangle , |xz,-\rangle\}$.  
Due to the symmetry properties of this manifold it only couples to $Q_2$ and $Q_6$. 
We have verified numerically that the coupling to $Q_6$ is too small to be relevant to the present discussion. In the sequel we therefor ignore the coupling to  $Q_6$. 
The effect of spin-orbit coupling is to mix these states and form a set of non-degenerate spin-orbital doublets. 
The electron-lattice interaction and the spin-orbit coupling at a given site $j$ are described by the Hamiltonian \cite{Abragam_Electron_1970,Eremin_alternating_2011}
\begin{eqnarray}\label{ham}
    H= \frac{P^2}{2M}+\frac{kQ^2}{2}+
    \frac{V}{2} \left[\hat{l}_{+}^2+\hat{l}_{-}^2\right] Q + 
     \lambda\hat{l}_{z}\hat{s}_{z}
\end{eqnarray}
Here $k$ is the vibrational force constant, $V$ is the electron-lattice coupling constants and $P$ the momentum of the vibrational coordinate. Expressing the electronic hamiltonian on the basis $d_{\pm 1, \sigma}=d_{xz, \sigma}\pm i d_{yz, \sigma}$  we obtain a blockwise diagonal matrix with 2x2 blocks
\begin{eqnarray}\label{ham1}
    H&=&\frac{P^2}{2M}+\frac{kQ^2}{2}+
    \begin{pmatrix}
     \lambda/2& i  V Q\\
     -i  V Q & -\lambda/2\\
    \end{pmatrix}. 
\end{eqnarray}
We have to take into account the inter-site coupling of the Jahn-Teller active sites. 
For the purpose of the present discussion we assume a strong coupling, implying uniform displacement of all $Q(j)$ corresponding to the different lattice sites $j$. 
We define the dimensionless parameters $v=2V^2/(k\lambda)$, $h= 4 H V^2/(k\lambda^2)$, $q=2 Q V/\lambda$, $p=P\lambda /(2\hbar V)$, $\mu=M k \lambda^4/(16\hbar^2 V^4)$, and obtain
\begin{eqnarray}\label{ham2}
    h&=&\frac{p^2}{2\mu}+\frac{q^2}{2}+
    v\begin{pmatrix}
     1& i  q\\
     -i  q & -1\\
    \end{pmatrix}. 
\end{eqnarray}
Straightforward diagonalization, and treating $q$ as a classical variable, constitutes the adiabatic potential
\begin{eqnarray}\label{uq}
    u(q)&=&\frac{q^2}{2}- v\sqrt{1+q^2} 
\end{eqnarray}
The first important observation is, that for $v<1$ the absolute minimum occurs at $q=0$, hence symmetry breaking requires that $v>1$. In the present material the spin-orbit parameter is $\lambda=-30$ meV\cite{Eremin_alternating_2011}. 
Furthermore, using experimentally-based data for RbMnF$_3$ \cite{Solomon_Comparison_1974}) and  KMgF$_3$\cite{muramatsu_method_1979,brik_electronphonon_2004}, we obtain $V=3^{1/2}V_E/2=-0.26\pm 0.05$ eV$/\AA$, and the relevant force constant is $k=2$eV$/\AA^2$. 
With these parameters the coupling constant $v=1.5 \pm 0.5$, which is above the critical value $v=1$. 
It is furthermore natural to suppose that $v$ has a temperature dependence due to thermal fluctuations, which limits the range where $v(T)>1$.  

Anti-ferro-magnetism orders the spins along the the c-axis. Due to the spin-orbit coupling this effectively suppresses the Jahn-Teller coupling. The minimal model describing this state of affairs is
\begin{eqnarray}\label{vmt}
   v(T,m)&=&1+\beta(1-\alpha m^2)(1-T/T_{c2})
\end{eqnarray}
where $\alpha $ is a constant describing the coupling between the Jahn-Teller order parameters, and the anti-ferromagnetic order parameter $m$. The coupled Jahn-Teller and anti-ferromagnetic order are described by the free energy
\begin{eqnarray}\label{gl1}
    u(q,m)&=&\frac{q^2}{2}- v(T,m)\sqrt{1+q^2} + f(m)
    \nonumber\\
    f(m)&=&f_2 m^2 +f_4{m^4}+ f_6 m^6 
\end{eqnarray}
The free energy minimum corresponds to the condition $du/dq=du/dm=0$. The Jahn-Teller transition at $T_{c2}$ (where in the case of Sr$_2$VO$_4$ there is no anti-ferromagnetic order) follows from the condition
\begin{eqnarray}\label{gl2}
    m(T_{c2})&=&0
    \nonumber\\
    v(T_{c2},0)&=&1
\end{eqnarray}
A second order magnetic transition is described by the set of parameters $f_6=0$ , $f_4=1/2$ and $f_2=\gamma (T/T_m-1)$. The anti-ferromagnetic transition occurs at $T_N$, where $T_N < T_m$ due to the competition between Jahn-Teller ordering and anti-ferromagnetism. Lowering the temperature below $T_N$, the anti-ferromagnetism progressively suppresses the Jan-Teller ordering. This process is completed at the temperature $T_{c1}$, which follows from the self-consistent relation
\begin{eqnarray}\label{gl3}
    v(T_{c1},m(T_{c1}))&=&1
\end{eqnarray}
For illustration we show in Fig. \ref{order}  the coupled order parameters $q$ and $m$  as a function of temperature using Eq. \ref{gl1}.  The input parameters $\alpha=0.35$, $\beta=1$, $\gamma=50$, $T_{c2}=136$ K, $T_m=105$ K, were chosen such as to obtain values for the transition temperatures of the coupled system $T_{N}$ (105 K), and $T_{c1}$ (100 K) close to the experimental ones. 
%
%
%

%

\end{document}